# SUPPORT VECTOR MACHINE-BASED FIRE OUTBREAK DETECTION SYSTEM

Uduak Umoh, Edward Udo and Nyoho Emmanuel

University of Uyo, Faculty of Science, Department of Computer Science, Uyo, Akwa Ibom State, Nigeria, PMB 1017

## ABSTRACT

*This study employed Support Vector Machine (SVM) in the classification and prediction of fire outbreak based on fire outbreak dataset captured from the Fire Outbreak Data Capture Device (FODCD). The fire outbreak data capture device (FODCD) used was developed to capture environmental parameters values used in this work. The FODCD device comprised DHT11 temperature sensor, MQ-2 smoke sensor, LM393 Flame sensor, and ESP8266 Wi-Fi module, connected to Arduino nano v3.0.board. 700 data point were captured using the FODCD device, with 60% of the dataset used for training while 20% was used for testing and validation respectively. The SVM model was evaluated using the True Positive Rate (TPR), False Positive Rate (FPR), Accuracy, Error Rate (ER), Precision, and Recall performance metrics. The performance results show that the SVM algorithm can predict cases of fire outbreak with an accuracy of 80% and a minimal error rate of 0.2%. This system was able to predict cases of fire outbreak with a higher degree of accuracy. It is indicated that the use of sensors to capture real world dataset, and machine learning algorithm such as support vector machine gives a better result to the problem of fire management.*

## 1. INTRODUCTION

Fire is a combustion process which results in light, smoke, heat, flame and various hazardous gases [1]. Heat, oxygen and fuel are the three major elements of fire. The nature of the fire depends on the proportion of each of these elements [2]. Even though fire has always been useful for promoting the development of human society and it is extensively used in lot of applications such as cooking, manufacturing process and other processes, when fire is out of control, i.e., fire hazard or fire outbreak, it can cause a serious effect to human life as well as property. Fire outbreak can also damage the ecological environment on a great level. Fire outbreak detection is the process of sensing of one or more phenomena resulting from fire such as smoke, heat, infrared light radiation or gas. Fire outbreak detection is usually done manually by visual observation but it is a hazardous job that can put the life of a human being in danger. Thus, in order to prevent fire outbreak and reduce losses due to fire outbreak, and for human safety, there is a great requirement to develop an intelligent system for fire outbreak detection. By putting this type of system to perform the fire outbreak detection task in fire-prone areas of a building, it can help in avoiding fire accidents and the loss of lives. Traditionally the fire detection is done using only smoke detectors which is less reliable technique and results in false alarms and these fire detectors are unable to respond quickly and in reliable manner in hazardous conditions [3]. Unlike the traditional fire outbreak detection devices, this system implements a sensor approach in predicting the fire outbreak. This system is based on a machine learning technique called support vector machine for the detection of fire outbreak.

Support vector machines are data-driven and nonlinear approaches used for pattern classification problems which do not incorporate problem domain knowledge [4] [5]. Support vector machines are supervised learning techniques, which are a relatively new class of learning machines [6]. Furthermore, an SVM not only effectively learns complex relationships without heuristic feature parameters but is also suited to the use of a limited trained data set. The main goal of SVM theory is to use kernel functions to map the training data into high-dimensional feature space. A hyper plane is then found in this feature space that maximizes the margin between categories. The constructed hyper plane can then be used as a basis for classifying vectors of unknown classification with the largest margin between classes.

## 2. RELATED WORKS

In [7]), a Fire detection based on vision sensor and support vector machines is proposed, a new vision sensor-based fire-detection method for an early-warning fire-monitoring system. First, candidate fire regions are detected using modified versions of the detection of moving regions and fire-colored pixels methods. Next, since fire regions generally have a higher luminance contrast than neighbouring regions, a luminance map is made and used to remove non-fire pixels. Thereafter, a temporal fire model with wavelet coefficients is created and applied to a two-class support vector machines (SVM) classifier with a radial basis function (RBF) kernel. The SVM classifier is then used for the final fire-pixel verification. Experimental results showed that the proposed approach was more robust to noise, such as smoke, and subtle differences between consecutive frames when compared with the other method.

In [8] an SVM based forest fire detection using static and dynamic features is proposed. It is an approach for automatic forest fire detection from video based on 3D point cloud of the collected sample fire pixels, Gaussian mixture model is built and helps segment some possible flame regions in single image. Then the new specific flame pattern is defined for forest, and three types of fire colours are labelled accordingly. With 11 static features including colour distributions, texture parameters and shape roundness, the static SVM classifier is trained and filters the segmented results. Using defined overlapping degree and varying degree, the remained candidate regions are matched among consecutive frames. Subsequently the variations of colour, texture, and roundness, and area, contour are computed, and then the average and the mean square deviation of them are obtained. Together with the flickering frequency from temporal wavelet based Fourier descriptors analysis of flame contour, 27 dynamic features are used to train the dynamic SVM classifier, which is applied for final decision. The approach was tested with dozens of video clips, and it can detect forest fire while also recognizing the fire like objects, such as red house, bright light and flying flag.

In [9] an intelligent fire detection and mitigation system is studied.  The study proposed an adaptive fusion method for fire detection based on three sensors - Flame sensor, Temperature sensor and Gas sensor. Arduino Uno with Flame sensor, Temperature sensor and Gas sensor are used as source of the input data. Arduino IDE was used for system implementation. Sometimes the actual fire can be taken as the deceptive fire because the system considered multiple numbers of sensors's input to determine the actual fire event. [10] On "Fire Detection Using Support Vector Machines (SVM)" proposes a robust approach to wildfire detection using support vector machine (SVM) algorithm. The result was compared with logistic regression. SVM proofs superior in detecting cases of wildfire.

A design and implementation of a fire and obstacle detection control system is investigated based on Arduino Uno, using fire sensor, temperature sensor, smoke sensor, and PIR motion sensor on automobile [11]. The system was implemented with Matlab and used for system development.

The microcontroller unit responds to the instructions sent by the Matlab R2014 software according to the necessity of the application as well as triggers the relay and fan upon critical situations.

Design of a fire detection algorithm based on Fuzzy Logic, using temperature sensor to estimate the direction of fire disaster is carried out [12] While the sensor nodes used has the ability to periodically collect ambient temperature, communicate with neighbor sensors, and store the neighbors' information such as Nodes' ID, and the Nodes' coordinate, Temperature sensor is not enough to detect fire outbreak. Sensor node could not know their exact location in the grid. [13] presents an intelligent Fire Monitoring and Warning System (FMWS) that is based on Fuzzy Logic to identify the true existence of dangerous fire and send alert to Fire Management System (FMS). The study discusses design and application of a Fuzzy Logic Fire Monitoring and Warning System that also sends an alert message using Global System for Mobile Communication (GSM) technology. One of the most interesting real world phenomena that can be monitored by WSN is indoor or outdoor fire.

In [14] a study explores two fuzzy logic approaches, with temporal characteristics, for monitoring and determining confidence of fire in order to optimize and reduce the number of rules that have to be checked to make the correct decisions. The study assumes that this reduction may lower sensor activities without relevant impact on quality of operation and extend battery life directly contributing the efficiency, robustness and cost effectiveness of sensing network. [15], implements fuzzy logic on the information collected using sensors by passing them on to the cluster head using event detection mechanism. Multiple sensors are used for detecting probability of fire as well as direction of fire. Each sensor node consists of multiple sensors that will sense temperature, humidity, light and CO density for calculating probability of fire and azimuth angle for calculating the direction of fire. [16], employs fuzzy logic in fire detection system, which is also multi sensor based. The driving force behind this was to develop an efficient but a simple and error free system for critical applications. [17], presents an approach for fire detection and estimation robots based on type-2 fuzzy logic system that utilizes measured temperature and light intensity to detect fires of various intensities at different distances.

In [18], experiment is performed using the prototype, fire detection devices with an IoT connection to speed up the monitoring of fire hotspots. The use of fuzzy logic minimises false warnings from fire detection devices. The prototype can be used as a medium of learning for high school students majoring in computer engineering and networking. [19], develops a WiFi ESP8266 Board Compatible with Arduino IDE. [20], study temperature measurement with a Thermistor and an Arduino. [21], develop an IoT-based intelligent modeling of smart home environment for fire prevention and safety [22], carried out forest fire detection using wireless sensor. [23] [24] carry out a review of wildfire detection using social media. [25], designs and implement forest-fires surveillance system based on wireless sensor networks for South Korea Mountains.

## 3. RESEARCH METHODOLOGY

In this paper, SVM-based fire outbreak detection system is developed and is applied to detect cases of fire outbreak via the use of temperature sensor, smoke sensor and flame sensor. These sensors are connected to an Arduino nano microcontroller unit. The Arduino microcontroller interfaces with the system peripherals (sensors) through the input/output (I/O) pins. Values sensed by these sensors are read by the Arduino board via a C program running on the Arduino's ATMega processor. The read parameter values are sent in a request parameter as a sequence of text via an ESP8266 Wi-Fi module to a web interface that captures, pre-processes, stores the data for use by the system. The stored dataset is visualized in order to descend their relationship. 700 data points are captured for the dataset with the use of FODCD. 60% of the data are used for training dataset while 20% of the dataset are employed for testing and validation. The conceptual architecture of the SVM-based fire outbreak detection system is presented in Figure 1.

## 3.1. ARCHITECTURE OF THE SYSTEM

Standard The conceptual architecture of the proposed system depicts the structure, components, and the relationship between each component of the system. The conceptual architecture comprises the following: (i) Sensors (ii) Arduino Microcontroller (iii) Wifi Module (iv) Data Receptor (v) Dataset (vi) Classifier (vii) Data Visualization. Three SVM-based fire outbreak detection input sensors are employed in this work for environmental data capturing as shown in Figure 2(a, b, c) respectively. They include – DHT11 Temperature sensor for temperature measurement, MQ-2 Smoke sensor for measurement of the amount of smoke gas in the air, and LM393 flame sensor to measure light intensity. These sensors provide a source of input values to the system where environmental parameters related to fire outbreak are captured. The study use Arduino Nano V3.0 Microcontroller Board with micro chip, adapted from [18] and presented in Figure 3, to interface all the system peripherals (sensors, and WI-FI module). The Arduino Microcontroller board serves as the brain of the fire outbreak data capture device. Its main function is to read parameter values from the connected sensors and send same to the classifier module via an ESP8266 WIFI module. The ESP8266 is used to add wireless network functionality to the fire outbreak data capture device (FODCD). The ESP8266 acts as a web server; hence its data can be viewed using a web interface programmed as an android application. The ESP8266 WIFI module used in this work is presented in Figure 4.

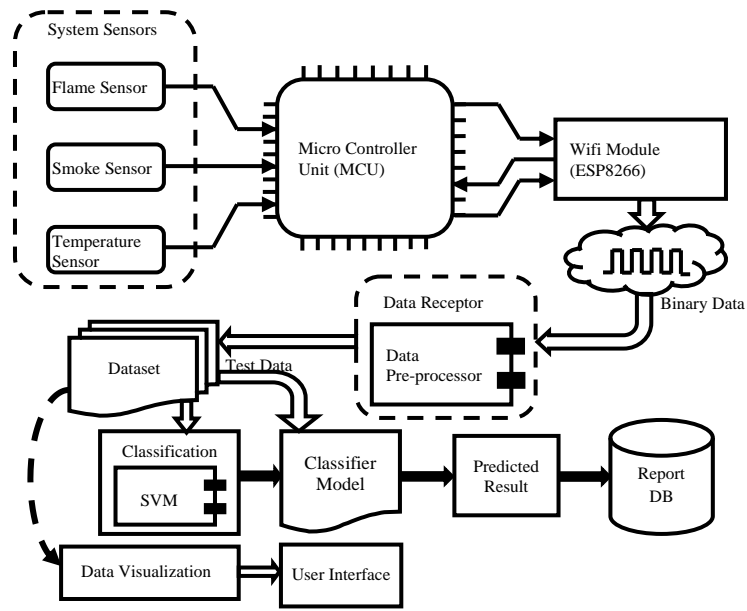

Figure1. The Conceptual Architecture of SVM-Based Fire Outbreak Detection System

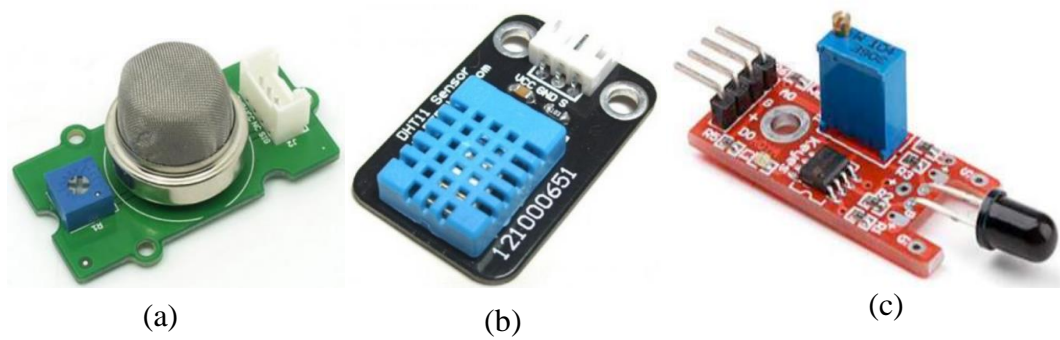

Figure 2: (a) MQ-2 Smoke Sensor (b) DHT11 Temperature Sensor (c) LM393 Flame Sensor employed in SVM-Based Fire Outbreak Detection

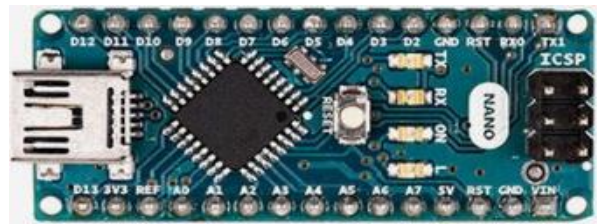

Figure 3. Arduino Nano V3.0 board for Interfacing Sensors and WI-FI module

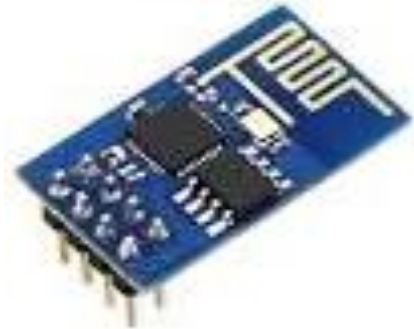

Figure 4: ESP8266 WIFI Module

The Data receptor is implemented as a software module. It sends an HTTP request for data to the ESP8266 WIFI module and accepts a response in the form of "HTTP Response Text". The Data Receptor reads and extracts the relevant parameters (i.e. temperature, smoke, and flame values) from the "HTTP Response Text" using the heuristic algorithm in Figure 5. Figure 6 gives the circuit diagram of the fire outbreak data capture device showing the hardware components, their pin connection, and the relationship of each component with the other as used in this paper.

```
algorithm Pre-process(msg):Arr[]
begin:
        Define temp, smoke, flame;
        Define dataSize N;
        Define indexOfAsteric indx1, indexOfAsh indx2;
        Define substringVariablesubStr;
        Define array Variable Arr[];

        //find the position of char *
        Loop for i = 0 to N
                Let x = charAt(i)
                If x == "*"
                        indx1 = i;
                        break;
                end if
        end for

        //find the index of char #
        Loop for j = 0 to N
                Let y = charAt(j)
                if y == "#"
                        Indx2 = j;
                        break;
                end if
        end for

        //extract data substring and split to an array
        if( indx1 > -1 AND indx2 > -1)
                subStr = subString(indx1, indx2);
                Arr[] = split(subStr, ",");
                return Arr[];
        end if
end
```

Figure 5. Heuristic algorithm for data reading and extraction of Input Parameter Values by Data Receptor

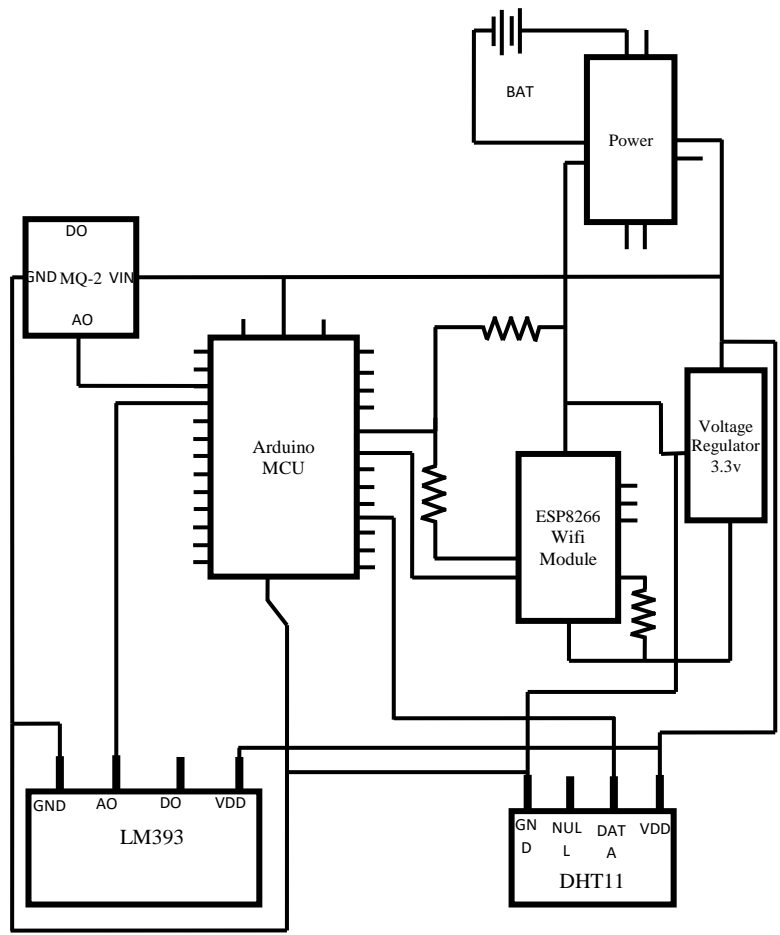

Figure 6. Circuit diagram of the fire outbreak data capture device

## 3.2. THE PROTOTYPE OF FIRE OUTBREAK DATA CAPTURE DEVICE

The FODCD device is developed for the purpose of extracting environmental parameters values related to fire outbreak. The FODCD provides the input parameter values used in testing the performance of this system. The breadboard is designed to connect the necessary components of the device together to help in testing and debugging of the device before the final soldering is done. The prototype of the fire outbreak data capture device is presented in Figure 7;

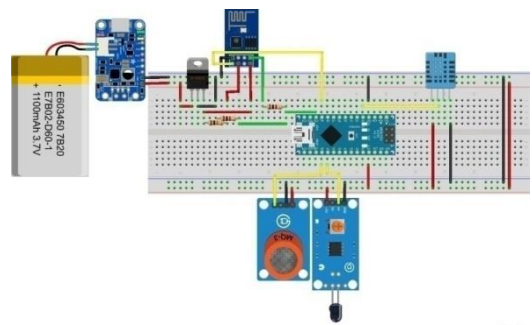

Figure 8. Prototype of the fire outbreak data capture device

### 3.3. FIRE OUTBREAK DATASET

The dataset used in this work is captured with the FODCD. The data collected includes the temperature, smoke, and flame parameter values. The dataset is made up of three features (temperature, smoke and flame) and one output label called "Fire Outbreak Detection". The label with values 0 is the negative class (i.e. "NoFireOutbreak") while 1 is the positive class (i.e. "FireOutbreak"). The sample of this data is presented in Table 1.

Table1. Fire Outbreak Collected Dataset

| Temp | Smoke | Flame | Label |
|------|-------|-------|-------|
| 25.977 | 50.529 | 464.37 | 0 |
| 36.301 | 60.853 | 464.47 | 0 |
| 18.554 | 36.632 | 359.63 | 0 |
| 29.317 | 52.735 | 387.89 | 0 |
| 16.351 | 30.961 | 234.26 | 0 |
| 19.748 | 38.247 | 283.3 | 0 |
| 25.229 | 47.722 | 471.83 | 0 |
| 33.063 | 54.949 | 593.49 | 0 |
| 34.735 | 66.18 | 574.92 | 0 |
| 25.041 | 42.024 | 412.61 | 0 |
| 15.946 | 27.564 | 208.08 | 0 |
| 28.005 | 55.083 | 450.8 | 0 |
| 42.141 | 74.721 | 796.43 | 0 |
| 17.825 | 30.052 | 336.47 | 0 |
| 17.423 | 32.546 | 312.48 | 0 |
| 15.365 | 28.35 | 211.6 | 0 |
| 30.165 | 56.058 | 558.45 | 0 |
| 42.137 | 81.57 | 753.23 | 1 |
| 33.566 | 65.797 | 483.05 | 0 |
| 37.682 | 66.021 | 516.66 | 0 |
| 45.647 | 79.257 | 920.62 | 1 |
| 15.358 | 27.124 | 277.24 | 0 |
| 19.993 | 35.956 | 301.45 | 0 |
| 35.196 | 57.753 | 604.37 | 0 |
| 27.631 | 47.253 | 409.88 | 0 |
| 41.154 | 73.372 | 666.96 | 0 |
| 32.217 | 53.757 | 528.32 | 0 |
| 45.072 | 77.857 | 848.01 | 1 |
| 19.699 | 31.73 | 323.23 | 0 |
| 24.379 | 44.86 | 311.9 | 0 |
| 16.406 | 29.229 | 257.16 | 0 |
| 48.478 | 94.916 | 826.49 | 1 |
| 48.24 | 88.208 | 845.21 | 1 |
| 16.552 | 30.01 | 218.67 | 0 |
| 38.36 | 64.808 | 503.67 | 0 |
| 39.138 | 70.889 | 791.07 | 0 |
| 45.98 | 84.589 | 822.88 | 1 |
| 49.955 | 86.207 | 872.83 | 1 |
| 24.913 | 46.242 | 374.64 | 0 |
| 36.601 | 58.797 | 621.71 | 0 |
| 32.489 | 56.149 | 572.02 | 0 |
| 15.163 | 30.077 | 221.83 | 0 |
| 34.102 | 62.306 | 627.75 | 0 |
| 32.325 | 55.507 | 410.25 | 0 |
| 18.398 | 34.357 | 329.47 | 0 |
| 45.018 | 73.837 | 757.65 | 1 |
| 24.17 | 43.099 | 494.4 | 0 |
| 28.059 | 49.878 | 421.45 | 0 |
| 46.949 | 77.574 | 610.23 | 1 |
| 43.309 | 82.043 | 575.21 | 1 |
| 39.149 | 64.331 | 663.74 | 0 |
| 20.091 | 33.917 | 326.08 | 0 |
| 15.362 | 28.662 | 305.08 | 0 |
| 34.936 | 67.158 | 455.75 | 0 |
| 33.721 | 66.066 | 595.37 | 0 |
| 16.054 | 30.052 | 315.77 | 0 |
| 20.264 | 37.739 | 376.55 | 0 |
| 45.774 | 90.27 | 670 | 1 |

### 3.4. FIRE OUTBREAK DATA VISUALIZATION

Data visualization are performed on the fire outbreak captured data where different graphs are plotted to visualize the collected dataset in order to get a better understanding of the data structure. The different tools used in the dataset visualization include Lag plot, Correlation graph, and Andrew curveline. The results are obtained based on each visualization tool and are displayed in section four of this paper.

### 3.5. SUPPORT VECTOR MACHINE FOR THE FIRE OUTBREAK DETECTION

Support Vector Machine (SVM) is used to classify the data points (temperature, smoke and flame) into a positive class "FireOutbreak" with label 1 or negative class "NoFireOutbreak" with label 0. The Input data (training data) are formulated as shown in Equation 1;

$$(x_1, y_1) \dots \dots (x_n, y_n) \tag{1}$$

Where:

$x$ is the feature set, $y$ is the label, $x_i = x_i^1, x_i^2, \dots \dots x_i^d$, Where:$x_i^j$: is a real value and $y_i = \{0, 1\}$ with 0 representing "No Fire Outbreak" and 1 representing "Fire Outbreak"

In this work the RBF (Radial Basis Function) Kernel is used to map the non-separable training data from input space to feature space in order to find an optimized hyper plane that correctly segregates the data. The RBF kernel is presented in Equations 2 and 3 respectively.

$$K(\overrightarrow{x_i}, \overrightarrow{x_j}) = \emptyset(\overrightarrow{x_i})^T \emptyset(\overrightarrow{x_i}) \tag{2}$$

$$K(\overrightarrow{x_i}, \overrightarrow{x_j}) = \exp(-\gamma \, ||x_i - x_j||^2) \tag{3}$$

Where: $\gamma$ is $\frac{1}{2\sigma^2} > 0$, $x_i$ is the support vector points, $x_j$ is the feature vector points in the transformed space and $K(\overrightarrow{x_i}, \overrightarrow{x_j})$ is the Kernel function. The kernel function calculates the dot product of the mapped data points in the transformed feature space. The optimal hyper plane that segregates between the two classes ("FireOutbreak", "NoFireOutbreak") is found using equation 4;

$$w^T.x + b = \sum_{i=1}^{l} \alpha_i \, y_i \, \emptyset(\overrightarrow{x_i})^T \emptyset(\overrightarrow{x_i}) + b = 0 \tag{4}$$

The classification frontiers are found using;

$$w \, \emptyset(x) + b = \, 1 \text{: for points labelled as "FireOutbreak"} \tag{5}$$

$$w \, \emptyset(x) + b = \, 0 \text{: for points labelled as "NoFireOutbreak"} \tag{6}$$

The optimal weight vector (w) is given by;

$$\overrightarrow{w} = \sum_{i=1}^{l} \alpha_i \, y_i \, \emptyset(\overrightarrow{x_i}) \tag{7}$$

This work uses the dual formulation of SVM algorithm presented as a maximization problem over $\alpha$ is given in Equation 8;

$$\max \sum_{i=1}^{l} \alpha_i - \frac{1}{2} \sum_{i=1}^{l} \sum_{j=1}^{l} \alpha_i \, \alpha_j \, y_i \, y_j \, \emptyset(\overrightarrow{x_i})^T \emptyset(\overrightarrow{x_i}) \qquad (8)$$

Subject to:

$$0 < \alpha < C \text{ and } \sum_{i=1}^{l} \alpha_i \, y_i = 0$$

Where: $\alpha_i$: is the weight vector, $y$: is the label vector, $\emptyset(\overrightarrow{x_i})^T \emptyset(\overrightarrow{x_i})$: is the kernel function, $C$: is the intercept. The decision function g used in making prediction is given as;

$$g(\vec{x}) = sgn(\vec{w}^T \vec{x} + b) \Rightarrow sgn(\sum_{i=1}^{l} \alpha_i \, y_i \, \emptyset(\overrightarrow{x_i})^T \emptyset(\overrightarrow{x_i}) + b) \qquad (9)$$

Where: $g(\vec{x})$: is the predicted label, $sgn$: is the sign of $(\vec{w}^T \vec{x} + b)$ (i.e. 1 or 0), $\alpha_i$: is the weight vector. The general steps taken to accomplish the classification and prediction of fire outbreak are given in Figure 9.

STEP 1: Load input training data
STEP 2: Separate data into $x_i$ (feature set) and $y_i$ (labels)
STEP 3: Map data from input space to feature space using RBF Kernel
STEP 4: Find an optimal hyper plane using Equation (6)
STEP 5: Find classification frontiers (support vectors)

Figure 9: General Steps for Classification and Prediction of Fire Outbreak

### 3.3. PERFORMANCE METRICS FOR FIRE OUTBREAK DETECTION

The True Positive Rate (TPR), False Positive Rate (FPR), Precision, Error rate, and accuracy performance metrics are employed to evaluate the accuracy of the developed fire outbreak system. The models are presented as follows;

$$TPR \, (Sensitivity \, or \, Recall) = \frac{TP}{TP+FN} \qquad (10)$$

$$FPR = \frac{FP}{FP+TN} \qquad (11)$$

$$Precision = \frac{TP}{TP+FP} \qquad (12)$$

$$Accuracy = \frac{TP+TN}{N} \qquad (13)$$

$$Error \, Rate = \frac{FP+FN}{N} \qquad (14).$$

## 4. RESULT AND DISCUSSION

This study employed Support Vector Machine (SVM) in the classification and prediction of fire outbreak based on fire outbreak dataset captured from the FODCD. The fire outbreak data capture device (FODCD) used was developed to capture environmental parameters in this work. The FODCD device comprised DHT11 temperature sensor, MQ-2 smoke sensor, LM393 Flame sensor, and ESP8266 Wi-Fi module, connected to Arduino nano v3.0.board. 700 data point were captured using the FODCD device, with 60% of the dataset used for training while 20% was used for testing and validation respectively. The SVM model was evaluated using the True Positive Rate (TPR), False Positive Rate (FPR), Accuracy, Error Rate (ER), Precision, and Recall performance metrics. Figure 10 shows the FODCD developed in this paper and used for capturing the input parameter (temperature, smoke and flame) values as shown in Table 1. Figure 11 gives the line graph indicating the levels of temperature values from Table 1. From the graph it is shown that the higher the line in the graph, the higher the value of temperature. Figure 12 presents the filled line graph showing the levels of the different smoke values received from the FODCD and used in training and testing of the system. The flame data is collected using FODCD and the graph of the result showing the different flame values is presented in Figure 13.

Figure 14 presents the Lag plot of the captured fire outbreak dataset used to check the randomness, suitability, outliers of the values in the dataset. Data is said to be random when there is no visible pattern in the lag plot. From Figure 14 (lag plot), we can observe a pattern in the plot which implies that our dataset is not random, hence the dataset is good for the machine learning task at hand. Figure 15 is the correlation graph used to visualize the relationship among the features and the relationship between the fire outbreak features and the labels (output variables). The correlation ranges from -1 to 1, with -1 meaning that the dataset is highly negatively correlated and 1 meaning that the dataset is highly positively correlated. From the Figure 15, the correlation between our features and the labels is about 0.75. This implies that the features are highly correlated to the predictor (class label). Figure 16 presents the Andrews curve used in visualizing multidimensional data by mapping each observation onto a function. The Andrew curve for the fire outbreak dataset shows the closeness of data points from one class to that of other class. In the plot, each colour used, represents a class (0 for negative class i.e. "NoFireOutbreak", 1 for positive class i.e. "FireOutbreak") and we can easily note the lines representing samples from the same class exhibit similar pattern. Figure 16 shows that there is a significant level of uniqueness between points that belong to the negative class ("no fire outbreak") and points that belong to the positive class ("fire outbreak").

Figure 17 gives the training result from the support vector machine for fire outbreak detection. It shows the segregation of the data points into their classes. The red points labeled 0 represents the negative class called "No Fire Outbreak" and the blue points labeled 1 represents the positive class called "Fire outbreak". The Circles are the support vectors (i.e. data points that participate in the deciding the segregating hyper plane). The figure shows a proper segregation of the data points into its various classes ("FireOutbreak" and "NoFireOutbreak"). Figure 18 shows the graph of correctly classified points during model testing. The blue points are the true positives (i.e. points that are truly predicted as "FireOutbreak") while the red points in the graph are true negative points (i.e. points that are truly predicted as "NoFireOutbreak"). Figure 19 shows the misclassified points. The red points in Figure 19 are points that are false negative points (i.e. points that are wrongly classified as "NoFireOutbreak").

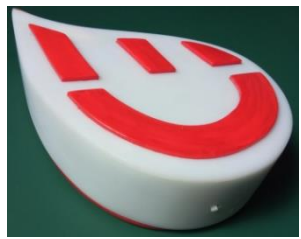

Figure: 10. Fire Outbreak Data Capture Device

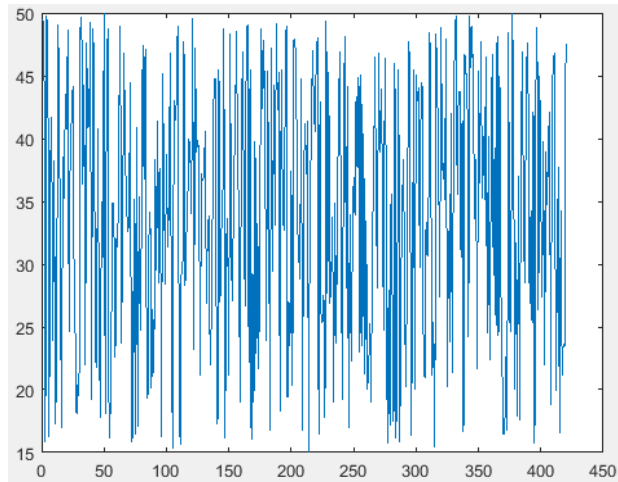

Figure 11. The Graph of Temperature Data Captured for Fire outbreak Detection

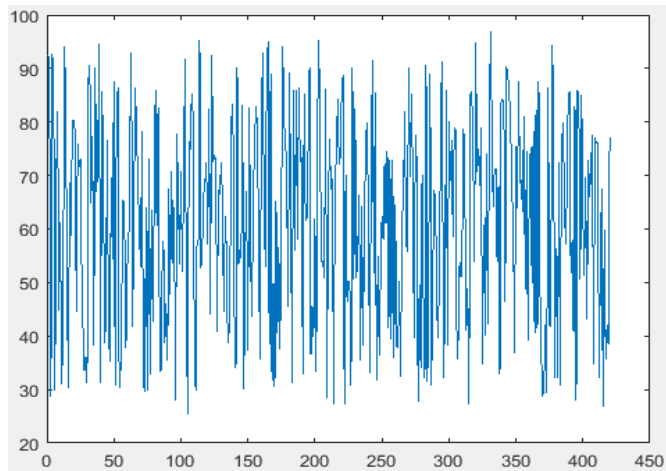

Figure12. The graph of the Captured Smoke Data for Fire outbreak Detection

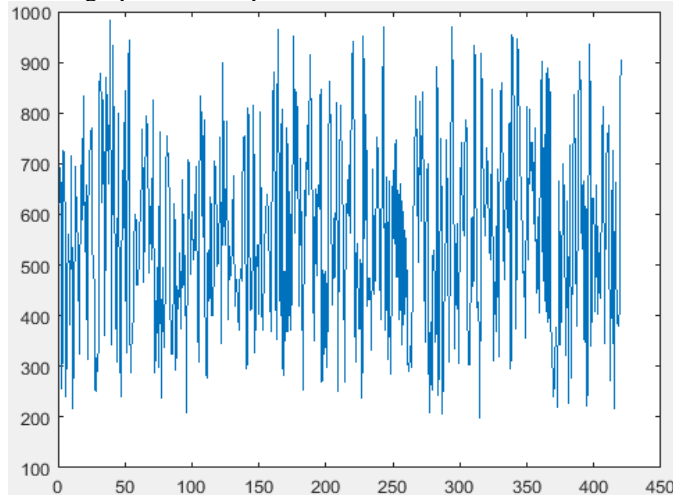

Figure 13. Graph of the Values of Flame Data Captured

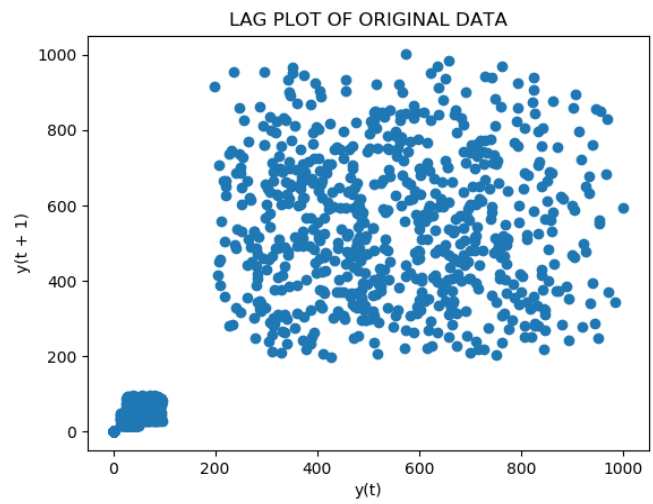

Figure 14. Lag Plot of the Fire Outbreak Dataset

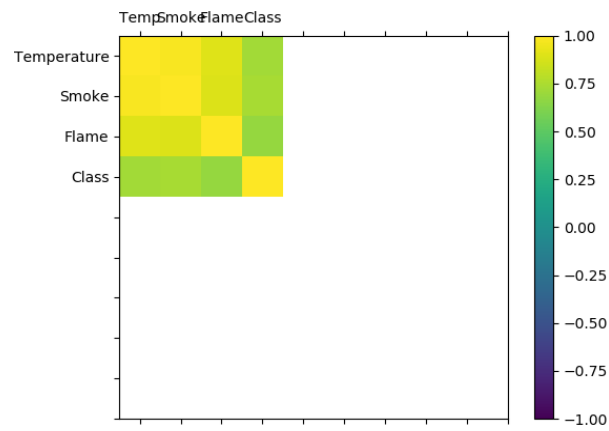

Figure 15. The graph of Correlation between the Fire the Outbreak Input and Output Variable

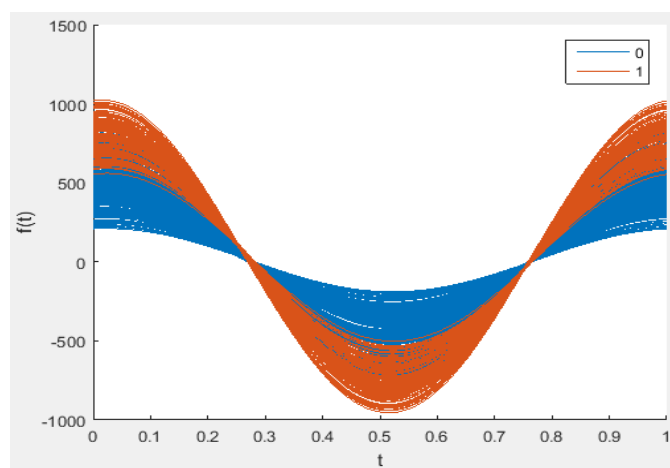

Figure 16. The Andrew Curve of the Fire Outbreak Dataset

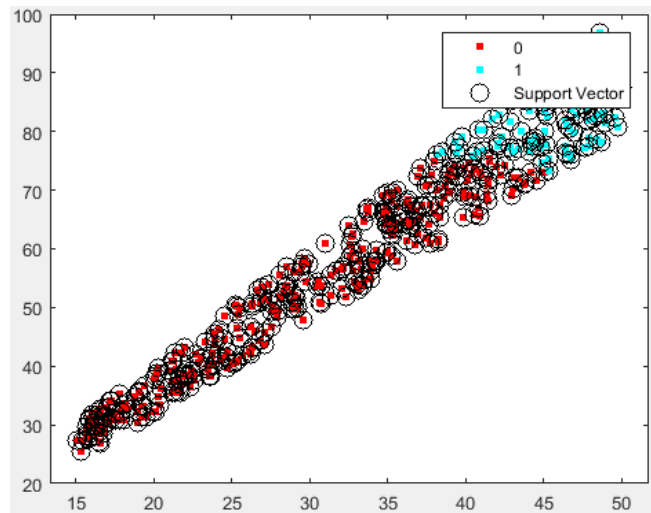

Figure 17. SVM Training result for Fire Outbreak Detection

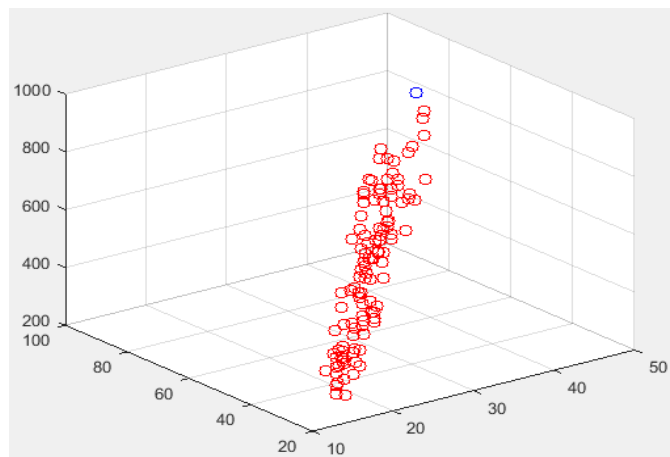

Figure 18. Correctly Classified Points

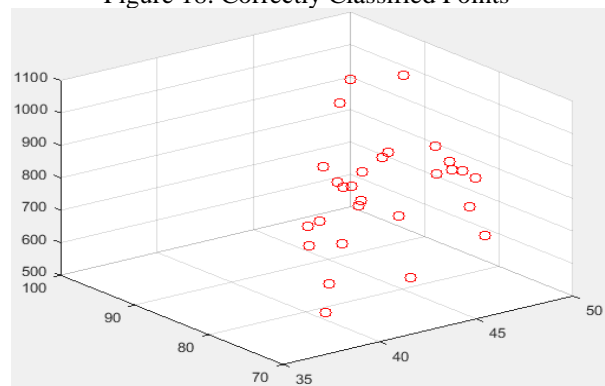

Figure 19. Misclassified points

The accuracy of the test result is evaluated using true positive, true negative, false positive, false negative, true positive rate, false positive rate, accuracy, precision and error rate. The test performance results of the model are presented in Table 2. From Table it is observed that, the values in the matrix represents true negative (i.e. 111), true positive (i.e. 1), false negative (i.e. 28), and false positive (i.e. 0). Figure 20 shows the Receiver Operating Characteristic (ROC) curve indicating the accuracy of the model on the test data. The accuracy rises when the curve is closer to the top left corner and deteriorates when the curve moves otherwise. From the ROC curve in Figure 20, the accuracy of 0.8 is attained from the model. The confusion matrix of the model testing is presented in Figure 21.

Table 2: Test Performance Result

| Metric | Symbol | Value |
|---|---|---|
| True Positive | TP | 1 |
| True Negative | TN | 111 |
| False Positive | FP | 0 |
| False Negative | FN | 28 |
| True Positive Rate (Recall) | TPR | 0.034483 |
| False Positive Rate | FPR | 0 |
| Accuracy | - | $0.8 \approx 80\%$ |
| Precision | - | 1 |
| Error Rate | - | 0.2 |

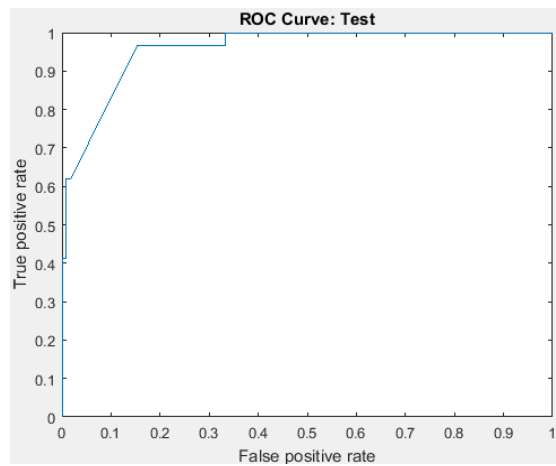

Figure 20. Test accuracy

| | 1 | 2 |
|---|---|---|
| 1 | 108 | 0 |
| 2 | 32 | 0 |

Figure 21. Confusion Matrix for testing result

## 5. CONCLUSION

This study employed Support Vector Machine (SVM) in the classification and prediction of fire outbreak based on fire outbreak dataset captured using Fire Outbreak Data Capture Device (FODCD). The fire outbreak data capture device (FODCD) is developed to capture environmental parameters values used in this work. The FODCD device comprised DHT11 temperature sensor, MQ-2 smoke sensor, LM393 Flame sensor, and ESP8266 Wi-Fi module, connected to Arduino nano v3.0.board. 700 data point were captured using the FODCD device, with 60% of the dataset used for training while 20% was used for testing and validation respectively. The results indicate that the support vector machine-based fire outbreak detection has been able to proffer solution to the problems associated with fire outbreak by providing continuous monitoring of environmental changes that are responsible for fire outbreak such as Temperature, Smoke and Flame. The system is also able to detect cases of fire outbreak by learning the parameters that could lead to fire outbreak. The system can predict a case of fire outbreak with accuracy of 80% with a minimal error rate of 0.2%. It is possible to detect and take immediate action during fire outbreak. When this is done, we can protect lives and properties as well as reduce the rate of depression resulting from fire disaster. The combination of sensors, and support vector machine learning algorithm gives a better result to the problem of fire management. In the future, more input parameters can be added and the system can be incorporated with deep learning tools for a better performance.

## Authors

**Umoh, U. A.** had received her Doctor of Philosophy (PhD) degree in Soft Computing from University of Port Harcourt,Rivers State, Nigeria in the year 2012, Master's degree in Database Management System from University of Port Harcourt, Rivers State, Nigeria in the year 2006 and Bachelor's degree from University of Uyo, Akwa Ibom State, Nigeria in 2007. She is currently working of as a Senior 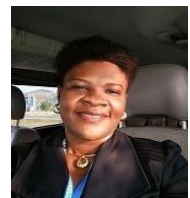 Lecturer, in the University of Uyo and  the Head of  Department  in the Department of Computer Science. She has published several articles in her areas in reputable national and international journals and has written some quality books in Computer discipline. Her area of interest include; Soft Computing (Fuzzy System, Neural network, Hybrid models), Database System, Data Communications, etc. She is a member of Nigerian Computer Society (NCS), Member, Computer Professionals Registration Council of Nigeria (CPN), Mathematical Association of Nigeria (MAN), Nigerian Women in Information Technology (NWIT), Universal Association of Computer and Electronics Engineers (UACEE), etc.

**Nyoho, E. E.** is a passionate software developer. He graduated from University of Uyo with Masters of Science degree in computer science. He has enormous experience in Software Development using Electron, Java, C, C++, Python, R. and Web Development using PHP, Java EE, JavaScript, HTML, CSS, Angular, Node.JS. He is also an android developer who has risen to the heights of publishing His applications on the Google Play Store. Due to his performance in software development, He was Awarded "Programmer of the year 2012" by National Association of Computer Science Students (NACOSS). He has carried out numerous projects in several areas of computing, including – Mobile Computing, Cryptography, Fuzzy Expert Systems, Image Enhancement, Steganography, Search, Machine Learning, and Embedded Systems.

**Udo, E. N.** BSc(Uyo); MSc(Port Harcourt),Ph.D(UNIBEN) He is a lecturer in the Department of Computer Science. He holds a Bachelor of Science degree in Computer Science from University of Uyo, Nigeria, a Master degree in Computer Science from University of Port Harcourt, Nigeria and a Ph.D degree from University of Benin, Nigeria. He holds a professional certificate in Networking (CISCO) and Systems 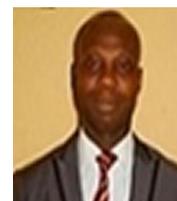 Engineering (Microsoft). He is internationally recognized CISCO Certified Academy Instructor. He is also certified as a Management Trainer by Centre for Management Development (CMD). He is a member of Nigeria Computer Society (NCS), a member of Computer Professionals Registration Council of Nigeria (CPN) and a member of IEEE (Computer section). He earned FGN/TETFUND Award in 2008 and 2012 for M.Sc. and Ph.D. Research respectively.

The authors declare that there is no conflict of interest regarding the publication of this paper.